\documentclass[journal]{IEEEtran}

\usepackage{amsmath}
\allowdisplaybreaks[4]
\usepackage{amsthm}
\usepackage{amssymb}
\usepackage{verbatim}
\usepackage{algorithm}
\usepackage{algorithmic}
\usepackage{bbding}
\usepackage{graphicx}
\usepackage{multirow}
\usepackage{multicol}
\usepackage{tabularx}
\usepackage{ragged2e}
\usepackage{booktabs}
\usepackage{makecell}
\usepackage{yhmath}
\usepackage{paralist}
\usepackage{color}
\usepackage{xr-hyper}
\usepackage[colorlinks,
 linkcolor=black,
 urlcolor=black,
 anchorcolor=black,
 citecolor=black]{hyperref}
\usepackage{cite}
\usepackage{enumitem}
\usepackage{xcolor,colortbl}
\usepackage{dsfont}
\usepackage{float}
\usepackage{epstopdf}
\usepackage{subfigure}
\usepackage{lipsum}

\begin{document}
\bstctlcite{BSTcontrol}

\title{
	Satellite-Terrestrial Integrated Fog Networks: Architecture, Technologies, and Challenges
}

\author{
	{
			Shuo~Yuan,
			Mugen~Peng,
			Yaohua~Sun
		}
	\thanks{
		Shuo Yuan, Mugen Peng (corresponding author), and Yaohua Sun are with the State Key Laboratory of Networking and Switching Technology, Beijing University of Posts and Telecommunications, Beijing 100876, China.
	}
}

\maketitle

\begin{abstract}
	In the evolution of sixth-generation (6G) mobile communication networks, satellite-terrestrial integrated networks emerge as a promising paradigm, characterized by their wide coverage and reliable transmission capabilities.
	By integrating with cloud-based terrestrial mobile communication networks, the limitations of low Earth orbit (LEO) satellites, such as insufficient onboard computing capabilities and limited inter-satellite link capacity, can be addressed.
	In addition, to efficiently respond to the diverse integrated tasks of communication, remote sensing, and navigation, LEO constellations need to be capable of autonomous networking.
	To this end, this article presents a satellite-terrestrial integrated fog network for 6G.
	Its system architecture and key technologies are introduced to achieve flexible collaboration between fog satellites and terrestrial cloud computing centers.
	In particular, key techniques with diverse challenges and their corresponding solutions are discussed, including integrated waveform design and resource management based on fog satellite onboard processing, as well as mobility management and native artificial intelligence based on cloud-fog collaboration.
	Finally, future challenges and open issues are outlined.
\end{abstract}

\IEEEpeerreviewmaketitle

\section{Introduction}
\label{sec:intro}

\IEEEPARstart{I}{ntegrating} low Earth orbit (LEO) satellite networks with terrestrial mobile communication networks and leveraging the complementary advantages of both to form satellite-terrestrial integrated networks (STINs) represent a key evolutionary direction for sixth-generation (6G) mobile communication systems \cite{wang2023road6}.
However, integrated tasks that include communication, remote sensing, and navigation, coupled with emerging intelligent services such as augmented/virtual reality and autonomous driving, pose significant challenges to the design of satellite payloads and networking methodologies within STINs \cite{zuo2024integratingcommunication}.
Traditional satellite payloads, due to their hardware being tightly coupled with task-specific functions, struggle to support multi-task operations (e.g., integrated communications and sensing).
They also lack efficient hardware support for inter-satellite and satellite-terrestrial collaboration.
For emerging intelligent services, there is a high demand for improved communication capabilities and advanced computing and sensing, which imposes additional requirements on satellite payload design.

To overcome these challenges, software-defined satellite payload and fog computing technologies have been introduced in STINs.
For example, the architecture of software-defined STINs has been explored in \cite{yuan2023softwaredefined} to improve the scalability and flexibility of network management.
However, these studies have not fully addressed the potential of fog computing within STINs or detailed satellite-based collaboration mechanisms.
Satellites with software-defined payloads can function as fog satellite nodes (FSNs), capable of on-demand payload reconfiguration for various tasks including communication and sensing, supported by onboard fog computing capabilities.
In this context, fog computing focuses on integrating signal processing and computing at the network edge, such as base stations or satellites, or on terminals for device-to-device communication and local computing \cite{peng2016fogcomputingbasedradio}.

In addition, the integration of FSNs with cloud-based terrestrial mobile networks offers a promising solution to overcome the inherent resource constraints of individual satellites.
This cloud-fog collaboration enables dynamic service-oriented reconfiguration of functions on demand across both FSNs and terrestrial nodes.
A prime example is electronic reconnaissance, where FSNs can be dynamically configured to perform radar signal processing, high-resolution imaging, and target detection, thereby significantly reducing data transmission volume.
Cloud computing centers then further conduct advanced post-analysis of the pre-processed data, substantially improving target identification accuracy.

\begin{figure*}[htbp]
	\centering
	\includegraphics[width=0.75\textwidth]{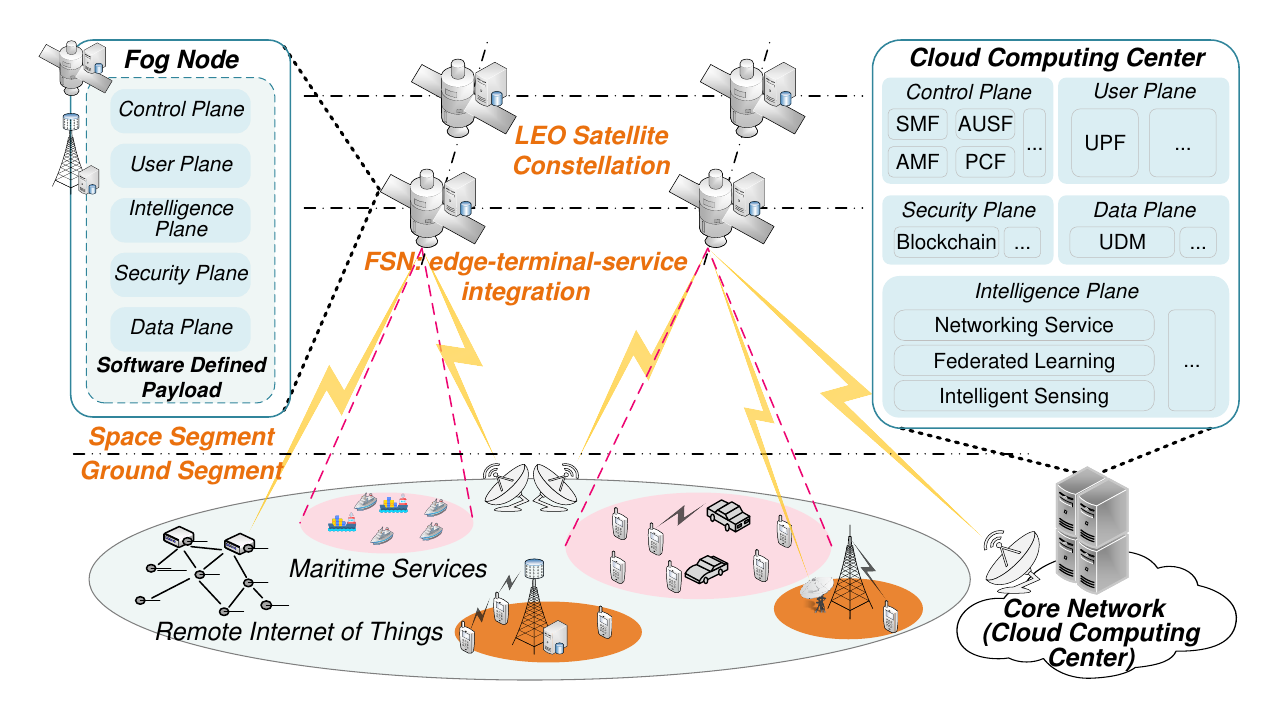}
	\caption{System architecture of STIFNs.}
	\label{fig:systemArchitecture}
\end{figure*}

However, this integration presents several challenges, including heterogeneous network architectures, complex resource allocation, increased energy consumption, and frequent handovers.
Moreover, the rapid growth of satellites, terminals, and services, coupled with the dynamic nature of STIN topologies \cite{yuan2024jointbeam}, makes traditional manual network management inefficient, resulting in significant capital and operational costs for network operators.
The remarkable progress in artificial intelligence (AI) has sparked extensive research into its capabilities for network management, which have been widely recognized.
In \cite{alhomssi2024artificialintelligence}, the applications of AI for massive satellite networks have been discussed, covering channel estimation, signal detection and demodulation, and access network optimization.
However, these studies do not address the benefits of fog computing and cloud-fog collaboration for AI, nor the advantages of distributed AI for autonomous networking.
Satellite-terrestrial integrated fog network (STIFN), an evolution of STINs that deeply incorporates fog computing principles, emerges as a promising paradigm to facilitate these essential collaborations, thus establishing native AI networks to address the challenges above effectively.

Motivated by these considerations, this article explores the system architectures and technological principles of STIFNs.
We propose a specific system architecture for STIFNs that delineates the collaboration between FSNs and cloud-based terrestrial communication networks, supported by two representative deployment cases.
The article examines critical STIFN technologies, including integrated waveform design and resource management leveraging fog satellite onboard processing, as well as mobility management and native AI capabilities enabled by cloud-fog collaboration, with corresponding performance analyses.
Future challenges and open issues are also discussed.

The remainder of this article is organized as follows.
The system architecture of STIFNs is introduced in detail in the following section.
Key technologies are discussed in the third section.
The future challenges and open issues are then highlighted, followed by our conclusion.

\section{System Architecture}

\subsection{Evolution from Fog Radio Access Network}
Heterogeneous cloud radio access networks (H-CRANs) \cite{peng2014heterogeneouscloud} decompose high power nodes into remote radio units and baseband units (BBUs), and consolidate the BBUs into a BBU pool for large-scale centralized cooperative signal processing and resource management.
Fog radio access networks (F-RANs) \cite{peng2016fogcomputingbasedradio}, evolving from H-CRANs, extend signal processing, computation, and storage functionalities to edge access nodes and terminals to form fog nodes, such as fog access points (FAPs) and fog user equipment (FUEs).
Consequently, with FAPs, F-RANs support localized processing and storage of user services, thereby establishing a cloud-fog collaborative radio access network.

However, the F-RAN architecture, tailored for terrestrial mobile networks, faces limitations in STINs due to the high dynamics of satellites, multi-layer network heterogeneity, and complex resource management requirements.
These challenges require a flexible and intelligent network architecture capable of handling frequent link state variations, orchestrating cross-domain resource allocation, and meeting diverse service requirements, which exceed the capabilities of F-RANs.
To bridge this gap, we propose a novel STIFN architecture for 6G, evolving from traditional F-RANs.
This architecture leverages terrestrial cloud computing and FSN-enabled edge-terminal-service integration to facilitate cloud-fog collaborative resource management and service provisioning.
The envisioned STIFN aims to deliver services with ultra-high data rates, ultra-low latency, ultra-high reliability, and ultra-high intelligence, supporting integrated tasks across communication, remote sensing, and navigation.

\subsection{STIFN System Architecture}

The proposed STIFN architecture is shown in Fig. \ref{fig:systemArchitecture}, which includes both space-based and ground-based segments.
The space segment consists of FSNs, which enable fog computing for diverse tasks and support inter-satellite links (ISLs) and satellite-terrestrial cooperation.
The ground segment includes terminals, base stations, ground gateways, and cloud computing centers, which offer substantial computing capabilities.

FSNs equipped with software-defined payloads represent a significant advancement over traditional single-task or single-service satellite payloads, enabling integrated communication, remote sensing, and navigation capabilities through dynamic payload scheduling and reconfiguration based on task requirements.
Current-generation satellites with early software-defined payloads, such as Eutelsat Quantum and TianZhi-2D, demonstrate advanced capabilities including mission-specific payload scheduling, on-demand beam resource allocation, and function updates on general-purpose computing platforms.
Meanwhile, research on onboard data processing is increasingly exploring edge graphics processing units, such as experimental hardware for the payload with Jetson TX2 (1 tera floating point operations per second (TFLOPS)) and Xavier NX (11 TFLOPS) \cite{zhang2022progresschallenges}.
Although computing power remains limited compared to terrestrial networks, these development trends are instrumental in unlocking the full potential of LEO satellites by virtualizing onboard computing and storage resources to support fog computing functionalities, including signal processing, AI computation, and edge caching.

By leveraging advanced technologies like software-defined networking and network functions virtualization, network and service functions can be designed and deployed across virtualized resources more flexibly, adapting to service demands and network conditions.
This flexible deployment of network functions forms the cornerstone of autonomous networking, facilitating scenario-specific implementation of networking paradigms. 
For example, in subnets that require intent-driven autonomous network management \cite{leivadeas2023surveyintentbased}, specialized network functions, including intent recognition, translation, and policy enforcement, can be deployed specifically to meet these requirements.

\begin{figure}[t]
	\centering
	\includegraphics[width=0.45\textwidth]{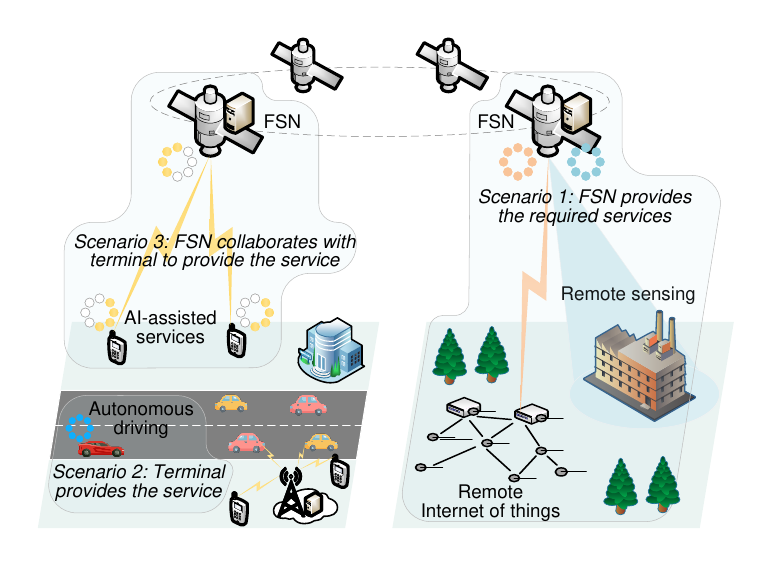}
	\caption{Three typical scenarios of FSN-enabled edge-terminal-service integration.}
	\label{fig:ETSintegration}
\end{figure}

\emph{\textbf{FSN-enabled Edge-Terminal-Service Integration:}}
With the increasing computing power and intelligence in satellites and terminals, FSNs enable edge-terminal-service integration for rapid service provisioning.
As depicted in Fig. \ref{fig:ETSintegration}, three primary scenarios of this integration are outlined:

\begin{itemize}
	\item \emph{Scenario 1:}
	      FSN caches and processes service data using on-orbit resources, reducing response latency and feeder link loads (e.g., onboard imaging and remote Internet of things).
	\item \emph{Scenario 2:}
	      Terminals collect, compute, and analyze service data locally, minimizing communication latency (e.g., localized vocabulary prediction of the input method and target detection of autonomous vehicles).
	\item \emph{Scenario 3:}
	      Terminals offload part of the service computation tasks to FSNs for processing. 
	      For services that support task partitioning, terminals collaborate with FSNs to provide services, while tasks involving private user data are processed locally to reduce the risk of data leakage.
\end{itemize}

\subsection{Collaboration Between FSN and Cloud-Based Terrestrial Communication Network}

FSNs can provide fog computing-based services to terminals and flexibly reconstruct communication and computing modes between satellites and terminals to accommodate varying service demands.
For example, to minimize latency, FSNs can perform service caching and edge AI computing directly.
To reduce backhaul traffic and optimize resource utilization, direct ISLs between FSNs can be established.
Furthermore, distributed collaborative computing between FSNs and terminals enables local data processing and rapid response to user requests.

As illustrated in Fig. \ref{fig:systemArchitecture}, a hybrid centralized-distributed service framework emerges, combining the capabilities of the cloud computing center and FSNs.
This framework facilitates cloud-fog collaborative resource allocation tailored to specific service scenarios and requirements.
For example, time-sensitive tasks can be offloaded to neighboring FSNs for edge computing, while wide-area and complex tasks can be addressed through inter-satellite collaboration within STIFNs.
Compute-intensive, latency-tolerant tasks can undergo preliminary processing on FSNs before being transferred to the terrestrial cloud computing center for more in-depth analysis.

\begin{figure*}[ht]
	\centering
	\subfigure[]{
		\includegraphics[width=0.4\linewidth]{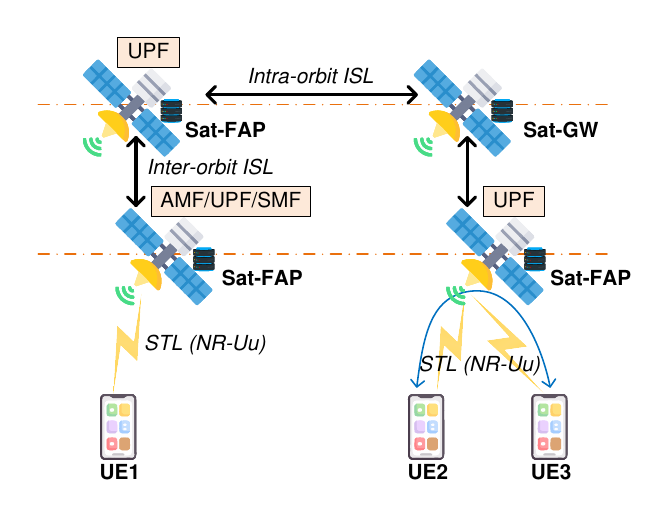}
		\label{fig:usecase1}
	}
	\hspace{1pt}
	\subfigure[]{
		\includegraphics[width=0.38\linewidth]{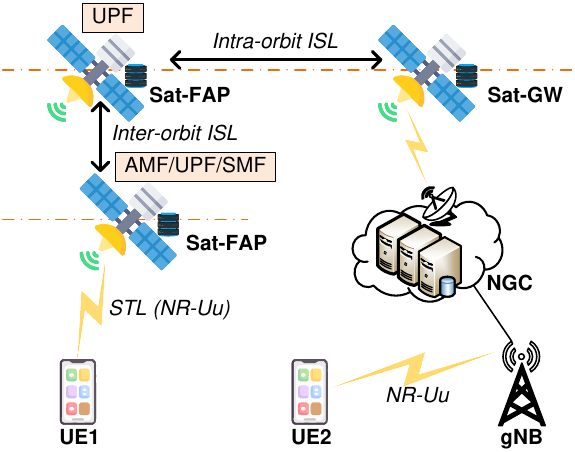}
		\label{fig:usecase2}
	}
	\caption{Two typical deployment cases: a) networking of multiple fog satellite nodes; b) cloud-fog collaboration in STIFNs.}
	\label{fig:usecase}
\end{figure*}

\subsection{Collaboration Between Five Planes}

The proposed STIFNs introduce a five-plane architecture, which includes the control plane, user plane, data plane, intelligence plane, and security plane, to enhance network management with adaptability, intelligence, and resilience.
As illustrated in Fig. \ref{fig:systemArchitecture}, this hybrid centralized and distributed framework enhances traditional approaches by providing additional planes dedicated to specific aspects of network functionality.

\begin{itemize}
	\item \emph{Control Plane:}
	      The control plane provides essential control functions for onboard signal processing, sensing, computation, storage, and security to meet the performance requirements of integrated tasks and emerging smart services. 
		  It enhances traditional network control functions, including the authentication server function (AUSF), policy control function (PCF), access and mobility management function (AMF), and service management function (SMF), while integrating essential control functions for onboard processing on FSNs.
	      These enhancements facilitate robust satellite-terrestrial cloud-fog collaboration, enabling effective mobility management and resource allocation.
	\item \emph{User Plane:}
	      The user plane is enhanced by improving the traditional user plane function (UPF) and providing programmable interfaces for intelligent networking strategies, information exchange, and service management.
	      This enhancement aims to improve the performance of the user plane protocol stack and increase efficiency and diversity in data processing and transmission.
	\item \emph{Data Plane:}
	      The data plane offers standardized and programmable data service interfaces that enable the structuring and storage of heterogeneous data from various sources at edge nodes or in the cloud.
	      Enhanced control plane functions related to user data management (UDM) and data service orchestration support dynamic management of network and service data such as distributed mobility management, endogenous AI services, edge caching, etc., thereby enabling full life-cycle data processing and management.
	\item \emph{Intelligence Plane:}
	      The intelligence plane serves as the cornerstone of native AI within the network, handling data analysis, decision-making support, and network self-optimization.
	      Leveraging machine learning, AI algorithms, and big data analytics, it can dynamically monitor network conditions, predict traffic patterns, and provide resource allocation optimization strategies in real time.
	      This plane fosters efficient collaboration between centralized AI in the cloud and distributed AI on fog nodes, addressing the increasing demands for intelligence in network management and service provisioning.
	\item \emph{Security Plane:}
	      The security plane manages the network’s security policies, including encryption, user authentication, threat detection, and intrusion prevention, forming a trusted security system based on proactive threat detection.
	      It also incorporates blockchain-based cross-domain data management to secure network operations and data transmission.
\end{itemize}

Through well-defined interfaces, each plane handles specific management tasks while supporting others. 
For example, while the control plane manages traffic routing centrally, the intelligent plane continuously monitors network conditions and suggests optimizations, reducing operational complexity. 
In addition, this virtualized function design, coupled with standardized interfaces, facilitates network scalability and upgrades.
For example, novel AI models, such as lightweight large language models, can be readily deployed as containers or functions on the intelligent plane to address increasingly complex AI demands, while the security plane can introduce new threat detection algorithms as needed.
By fostering efficient collaboration among the five planes, this network is expected to deliver services that meet diverse performance requirements, effectively realizing the 6G vision of cross-domain integration, in-network computing, and ubiquitous intelligence.

\subsection{Deployment Cases}

To illustrate the flexible networking types supported by STIFNs, we present two typical cases in Fig. \ref{fig:usecase}, which involve key components such as user equipment (UE), next generation node B (gNB), FSN, and next-generation core (NGC).

\emph{\textbf{Case 1: Networking of multiple fog satellite nodes.}}
FSNs, specifically designated as satellite fog access points (Sat-FAPs) to differentiate them from satellite gateways (Sat-GWs), are characterized by their onboard fog computing capabilities.
These FSNs can autonomously establish both intra-orbit and inter-orbit ISLs, enabling the formation of a self-organizing satellite network, as illustrated in Fig. \ref{fig:usecase1}. 
This deployment case is particularly suited for localized, multi-task scenarios, including remote Internet of things applications and emergency search and rescue operations.
FSNs can dynamically construct ISLs based on the evolving LEO satellite network topology and enable inter-satellite data exchange via Sat-GWs.
Moreover, FSNs can adaptively split, migrate, and execute the computing task among FSNs and terminals, realizing edge-terminal-service integration and implementing task-centric multi-satellite resource adaptive scheduling to accelerate service response times.
Terminals lacking direct satellite access can connect to terrestrial base stations supporting satellite-terrestrial link (STL) via the new radio Uu (NR-Uu) interface, thereby becoming part of an FSN-based self-organizing network.

\emph{\textbf{Case 2: Cloud-fog collaboration in STIFNs.}}
Due to limited computing and energy resources, FSNs may struggle to independently support compute-intensive tasks.
To address this issue, FSNs can collaborate with cloud-based terrestrial mobile communication networks within STIFNs, as depicted in Fig. \ref{fig:usecase2}.
This cloud-fog collaboration enables the efficient execution of complex tasks, such as high-precision target identification and wide-area meteorological monitoring, which rely on integrated communication and remote sensing.
For example, in an STIFN, telemetry data undergoes initial preprocessing at FSNs. 
The preprocessed results are subsequently transmitted to terrestrial cloud computing centers for more detailed telemetry target analysis and identification, facilitating cloud-fog collaborative data relay computing.
Furthermore, FSNs equipped with lightweight core network functionalities, such as AMF and SMF, can collaborate with the mobility management entity of terrestrial mobile communication systems to achieve cloud-fog collaborative handover decisions.
This collaboration facilitates seamless handover for users between inter-satellite and satellite-terrestrial connections.

\begin{table*}[ht]
	\scriptsize
	\centering
	
	\caption[Candidate Waveforms]{Candidate Waveforms for LEO Satellite Communications Compared to OFDM}
	\label{table:literatureReview}
	\begin{tabular}{p{1cm}p{3cm}p{4.3cm}p{4cm}}
		\toprule
		\textbf{Waveform}                                                                  & \textbf{Characteristics} & \textbf{Advantages} & \textbf{Disadvantages} \\
		\midrule
		OTFS                                                                               &
		Modulates signals in the {delay-Doppler domain}                             &
		1) Robust to multipath and Doppler effects; 2) high diversity; 3) low PAPR         &
		1) High implementation complexity; 2) high OOBE with rectangular pulse                                                                                       \\
		\midrule
		ODDM                                                                               &
		Orthogonal multiplexing in the {delay-Doppler domain}                       &
		1) Robust to multipath and Doppler effects; 2) high diversity; 3) low OOBE &
		High implementation complexity                                                  \\
		\midrule
		AFDM                                                                               &
		Multicarrier modulation using {affine transformation}                       &
		1) Robust to Doppler shifts; 2) full diversity in doubly dispersive channels       &
		High implementation complexity                                              \\
		\midrule
		GFDM                                                                               &
		{Block-based non-orthogonal multiplexing}                                   &
		1) Low OOBE; 2) highly flexible in time and frequency domains                      &
		1) High complexity in managing inter-symbol interference and inter-carrier interference; 2) high PAPR                                                                                                     \\
		\bottomrule
	\end{tabular}
\end{table*}

\section{Key Technologies}
\subsection{Integrated Waveform Design Based on Fog Satellite Onboard Processing}

The integration of communication, remote sensing, and navigation through integrated waveform design based on fog satellite onboard processing has garnered considerable attention due to notable improvements in diverse metrics such as frequency and energy efficiency \cite{yin2024integratedsensing}.
For example, the integrated sensing and communication waveform design aims to utilize shared hardware and spectrum resources for both communication and sensing tasks on the same platform using the same waveform, thereby reducing interference and power consumption compared to dual-system transmission modes adopted by radar and communication systems.

Traditional orthogonal frequency division multiplexing (OFDM) faces challenges in LEO satellite communications due to Doppler shifts and delay spread.
Several alternative waveforms, as summarized in Table \ref{table:literatureReview}, have emerged as promising candidates for high-mobility STIFNs.
Specifically, orthogonal time frequency space (OTFS) modulates data in the delay-Doppler domain, offering high diversity by spreading symbols across time-frequency and low peak-to-average power ratio (PAPR) with strong Doppler resilience.
In \cite{liu2024amplitudebarycenter}, a novel OTFS transceiver architecture for integrated sensing and communication is presented, enabling dual-functionality for both sensing and communication via advanced signal processing techniques.
Orthogonal delay-Doppler multiplexing (ODDM), an evolution of OTFS, uses square-root Nyquist pulses to reduce out-of-band emission (OOBE) and improve spectral efficiency.
Affine Fourier division multiplexing (AFDM) applies chirp-based modulation through the affine Fourier transform, offering full diversity in doubly dispersive channels.
Generalized frequency division multiplexing (GFDM) provides flexible resource allocation using non-orthogonal subcarriers and pulse shaping, benefiting spectral efficiency despite potential interference challenges.
While each waveform offers advantages in specific scenarios, they also present unique implementation complexities. 
A detailed comparison of these waveforms is provided in Table \ref{table:literatureReview}, highlighting their potential for STIFNs and the trade-offs involved.

\subsection{Resource Management Based on Fog Satellite Onboard Processing}

Traditional centralized resource management often leads to prolonged response times and uneven load distribution across network nodes.
In contrast, fog satellite onboard processing enables rapid on-demand resource allocation and adaptive service scheduling, enhancing resource utilization and service quality.
Specifically, within FSNs, dynamic decisions can be made regarding service scheduling, such as service splitting and offloading, and resource allocation, including beam scheduling and cloud-fog collaborative computing.
For example, computing tasks like target recognition and obstacle avoidance can be executed locally or offloaded to the nearest fog nodes based on terminal performance.
On the other hand, complex tasks such as cross-domain objective tracking and real-time path planning, which entail high latency and computational requirements, can be handled through an inter-satellite or satellite-terrestrial collaborative resource allocation strategy within an STIFN to balance latency, energy efficiency, and reliability.

\begin{figure}[ht]
	\centering
	\includegraphics[width=0.45\textwidth]{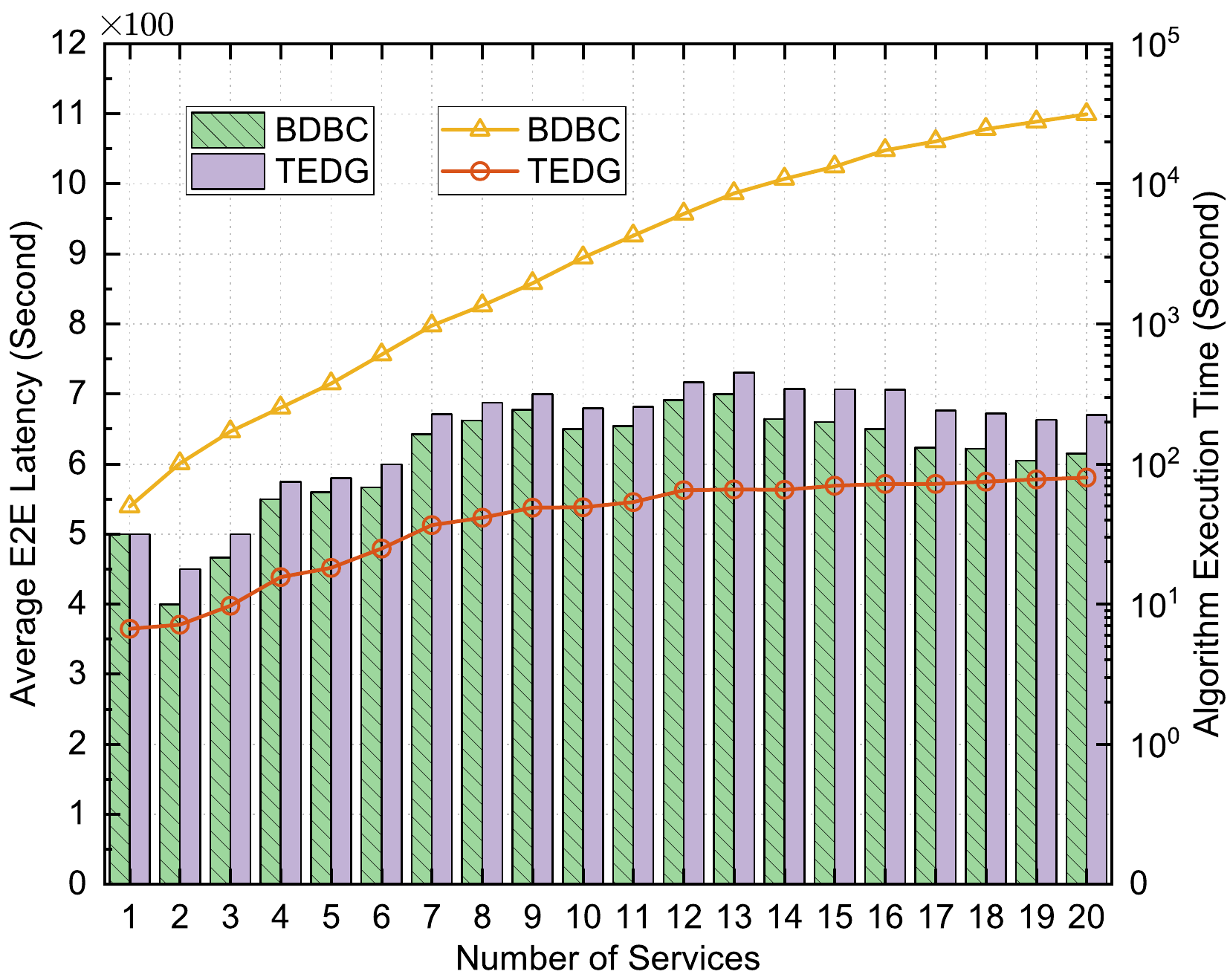}
	\caption{Service latency performance and computation complexity of Benders decomposition-based branch-and-cut (BDBC) algorithm and time expansion-based decoupled greedy (TEDG) algorithm. The bars indicate average service latency and the lines indicate algorithm execution time.}
	\label{fig:resourceAllocation}
\end{figure}

With advanced virtualization technologies, heterogeneous resources such as communication, computation, storage, and sensing can be represented in a unified form, which facilitates the joint allocation of multidimensional resources in STIFNs.
Current resource management approaches exhibit distinct trade-offs between performance and complexity. 
Convex optimization methods offer mathematical guarantees but face limitations in problem type and computational complexity, while matching theory and swarm intelligence-based algorithms provide rapid solutions but lack robust performance guarantees.
Our previous work \cite{yuan2024jointnetwork} contributes to this field with two complementary algorithms.
The Benders decomposition-based branch-and-cut algorithm optimizes computing function placement and data transmission paths in a satellite-terrestrial collaborative computing scenario with global optimality, while the time expansion-based decoupled greedy algorithm offers a low-complexity alternative maintaining solution quality, as illustrated in Fig. \ref{fig:resourceAllocation}.
Moreover, AI-based solutions, particularly deep reinforcement learning, are gaining attention for their adaptability to dynamic network environments.
As experimental hardware for onboard computing reaches TFLOPS-level processing power, more sophisticated and adaptive satellite-terrestrial resource management strategies are becoming feasible, paving the way for intelligent and responsive network operations.

\subsection{Mobility Management Based on Cloud-Fog Collaboration}

Due to the dynamic nature of LEO satellite networks, coverage areas and connections with terminals or gateway stations change frequently, leading to frequent handovers of user and feeder links.
Traditional handover mechanisms and mobility management methods face challenges such as complex tracking area management, inaccurate measurements, and signaling storms.
To address these challenges, cloud computing centers must consider satellite and terrestrial network characteristics, satellite ephemeris, user distribution, and service demands to intelligently adjust tracking area size and location.
This approach reduces tracking area updates and signaling overhead.
Collaborating with FSNs to optimize handovers between beams, satellites, and satellite-terrestrial connections is expected to enhance handover performance.

Many existing handover mechanisms typically focus on single metrics such as signal strength, coverage time, or node load.
However, in STIFNs, the channel characteristics are far more complex, involving severe satellite-terrestrial channel fading and Doppler shifts due to high satellite mobility.
To address these channel impairments, advanced link quality estimation technologies can be incorporated into the handover mechanism, enabling comprehensive channel state assessment that accounts for both large-scale fading (e.g., path loss and atmospheric effects) and small-scale fading effects, while implementing real-time Doppler shift compensation using frequency tracking algorithms \cite{zhu2024timingadvance}.
Furthermore, a holistic handover mechanism that incorporates multiple decision metrics, including communication link quality, satellite load balancing, and handover frequency and latency, can effectively reduce both the frequency and failure rates of handovers.
In addition, satellites with onboard gNB functions can use ISLs via the Xn interface to create more handover opportunities.
Compared to handovers conducted via the NG interface, ISL-enabled handovers through the Xn interface can significantly decrease the interactions between feeder links and terrestrial AMF elements, thereby reducing handover latency and feeder link overhead.

\subsection{Native AI Based on Cloud-Fog Collaboration}

Centralized cloud AI, with its abundant resources, offers high-speed, high-quality intelligent services for handling large-scale data and complex computing tasks, such as video analysis and recommendation systems.
From a network management perspective, it enables global network optimization, including long-term trend analysis of network traffic, coarse-grained resource scheduling strategies, and monitoring and mitigating security threats.
In contrast, distributed fog AI, deployed on resource-constrained nodes like fog satellites, engages in lightweight real-time data processing and local decision-making.
For example, in intelligent traffic systems, fog AI can dynamically adjust traffic light control strategies based on real-time traffic conditions and vehicle access requests.

\begin{figure}[htbp]
	\centering
	\includegraphics[width=0.48\textwidth]{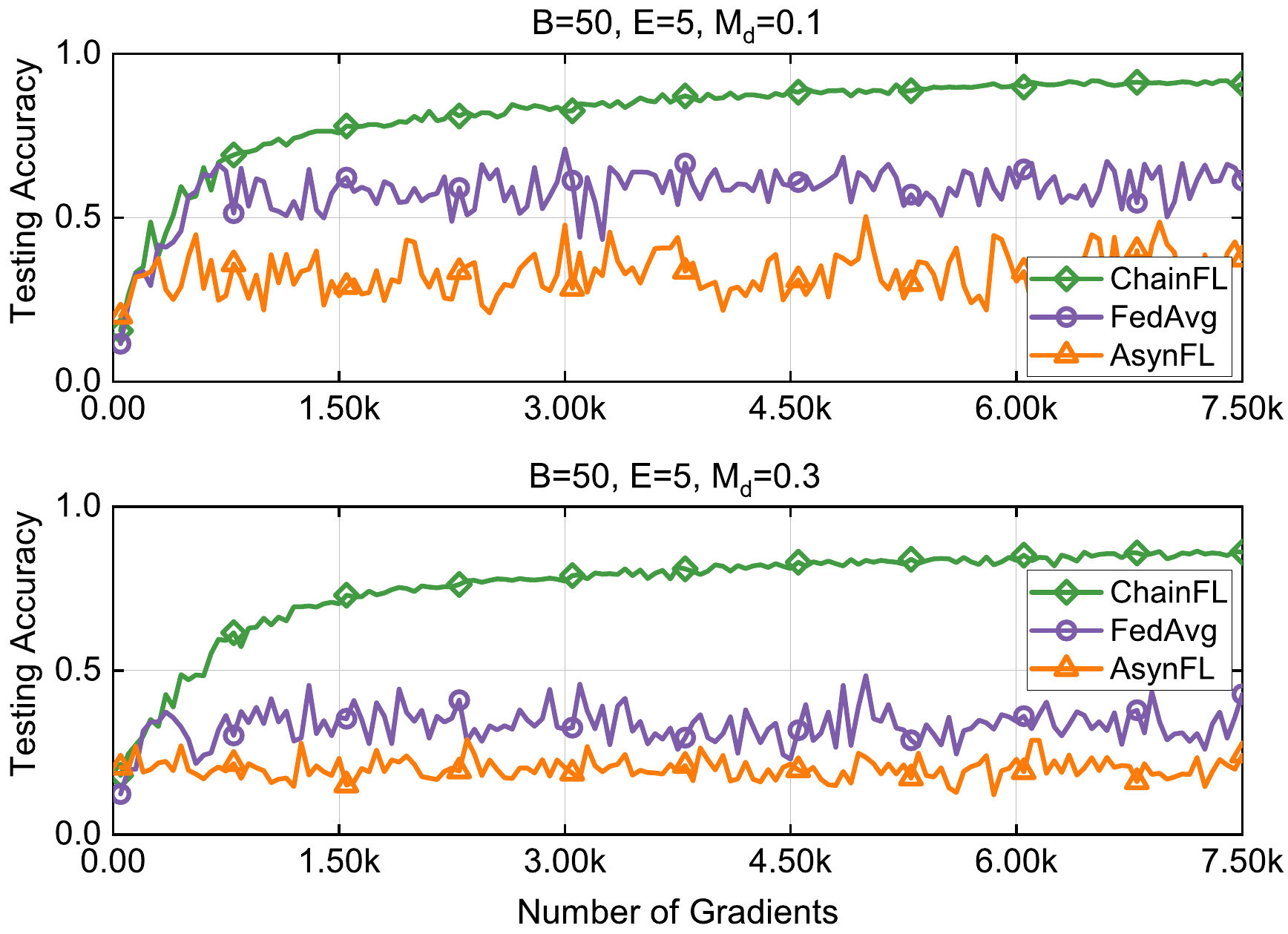}
	\caption{Robustness of distributed AI frameworks under different malicious device ratios (Mini-batch size (B), number of local epochs (E), malicious device ratio ($M_{d}$)).}
	\label{fig:federatedLearning}
\end{figure}

Efficient collaboration between centralized cloud AI and distributed fog AI is crucial for enabling native AI networks, significantly enhancing the intelligence of service delivery and network management.
To facilitate secure and efficient data exchange between cloud and fog nodes, the authors of \cite{yuan2024secureefficient} propose ChainFL, a blockchain-based federated learning architecture for distributed networks.
ChainFL enables comprehensive network analysis and optimization using network-wide data without centralized data collection and storage, thereby enhancing data privacy, reducing backbone network traffic, and alleviating the burden on cloud computing centers.
As shown in Fig. \ref{fig:federatedLearning}, compared to typical federated learning training mechanisms FedAvg and AsynFL, ChainFL exhibits significantly improved accuracy under scenarios where 10 percent and 30 percent of devices in the network are malicious.
By merging the benefits of blockchain and federated learning, this cloud-fog AI collaboration promotes model generalization and boosts network intelligence, offering a secure solution for cloud-fog AI collaboration in STIFNs.

\section{Challenging Work and Open Issues}

\subsection{Satellite-Terrestrial Integrated Network Slicing}

For remote private networks serving vertical industry customers with differentiated network service demands, end-to-end network slicing enabled by satellite-terrestrial integration is expected to provide highly customized services with assured quality, security, and isolation.
By scheduling and deploying network functions on demand, such as network self-optimization module in the intelligent plane and routing planning module in the control plane, and implementing distributed self-organizing networks, satellite-terrestrial integrated network slicing can be achieved more efficiently.
In addition, designing an end-to-end slice quality of service assurance mechanism across terminals, satellites, and terrestrial networks to meet the requirements of the slice service level agreement contributes to unified network slicing management and orchestration in STIFNs.

\subsection{Programmable Data Plane}

The expanding scale of satellite constellations and the rapid growth in communication traffic present significant challenges for traffic control in STIFNs. 
In mega-constellations, traditional data planes struggle to quickly adjust forwarding strategies to accommodate frequent LEO satellite handovers and fluctuating link quality, resulting in congestion and load imbalances.
The rise of programmable devices enables the creation of programmable data planes, allowing STIFNs to dynamically adapt data forwarding based on real-time link conditions and satellite positions, and then supporting distributed routing strategies, enhancing responsiveness and improving load balancing and congestion control \cite{feng2024distributedsatelliteterrestrial}.

\subsection{Security Assurance}

Given the distributed nature of STIFNs, managing multiple heterogeneous devices and implementing distributed AI learning present significant security challenges.
Sensitive user information, such as user location and authorization information, can be vulnerable to data breaches through multi-hop connections in diverse networks, including untrusted satellite or terrestrial nodes.
Therefore, it is crucial to investigate secure cross-domain data storage and exchange methods.
Utilizing technologies such as blockchain-based data management and advanced encryption can enhance the security and integrity of network management and user service data during transmission.

\subsection{Hardware Experimentation}

To demonstrate the concepts and technologies of STIFNs, hardware experimentation is crucial, and multiple prototypes have been developed and tested.
For example, China Mobile developed the ``XingHe'' prototype for satellite-based core networks.
This prototype features plug-and-play hardware to support function reconfiguration for different task scenarios, along with Internet protocol-based software for on-orbit UPF reconfigurable deployment, and supports onboard computing.
To reduce reliance on fiber optic connections in remote industrial control, our team partnered with Yinhe Hangtian to experiment with networking two private networks via LEO broadband satellites, achieving high uplink data rates, low end-to-end latency, and immersive remote robot control.
With the advancement of technologies such as software-defined radio and inter-satellite laser communication, it is anticipated that reconfigurable fog satellite hardware platforms can be developed in orbit, paving the way for the formation of satellite-terrestrial fog networks.

\subsection{Standardization Work}

The standardization of STIFNs-related has progressed significantly through the 3rd Generation Partnership Project, from initial studies in Release 15 to a series of specifications in Releases 17 and 18. 
The RAN working group laid the foundational research through TR 38.811 and TR 38.821, defining satellite access deployment scenarios and channel models. 
These efforts led to specifications TS 38.108 and TS 38.101 series, covering radio frequency performance requirements of satellite access nodes and terminals. 
The system architecture integrated satellite access is defined in TS 23.501, with service requirements and roaming standards detailed in TS 22.261. 
Release 19 will mark key advancements with Phase 3 satellite access integration, Phase 2 security and management, and Stage 2 application enablement for AI/machine learning.
Complementing these developments, the zero-touch network \& service management (ZSM) committee of the European Telecommunications Standards Institute has standardized network slicing (GS ZSM 003), cross-domain orchestration (GS ZSM 008), and closed-loop automation (GS ZSM 009-1), collectively enabling the autonomous networking capabilities essential for STIFNs deployment. 
Looking ahead, further enhancements in areas like AI/machine learning management, computation integration, and end-to-end orchestration are expected to drive the continued advancement of STIFN-related standardization.

\section{Conclusion}
\label{sec:conclusion}

In this article, we have introduced a satellite-terrestrial integrated fog network architecture for 6G systems, which integrates cloud-fog collaborative computing into network management and service provisioning.
To fully explore the performance potential of satellite-terrestrial integrated networks, we have presented key technologies including integrated waveform design and resource management based on fog satellite onboard processing, as well as mobility management and native AI based on cloud-fog collaboration.
Within these four key techniques, we have summarized the diverse challenges and corresponding solutions.
Finally, we have discussed the challenging work and open issues that need to be investigated further.

\section*{Acknowledgments}
This work was supported in part by the National Key Research and Development Program of China under Grant 2021YFB2900200, in part by the National Natural Science Foundation of China under Grants 61925101 and 62371071, and in part by the Young Elite Scientists Sponsorship Program by the China Association for Science and Technology under Grant 2021QNRC001.

\appendices

\ifCLASSOPTIONcaptionsoff
	\newpage
\fi

\bibliographystyle{IEEEtran}
\bibliography{bibUsedinPaper,bstControlForIEEEtran}

\section*{Biographies}
\begin{IEEEbiographynophoto}
	{\textsc{Shuo Yuan}} [M] (yuanshuo@bupt.edu.cn) received his Ph.D. degree in information and communication engineering from Beijing University of Posts and Telecommunications (BUPT), Beijing, China, in 2024.
	In 2024, he joined BUPT, where he is currently a Postdoctoral Fellow.
	His research interests include LEO satellite communications and edge intelligence.
\end{IEEEbiographynophoto}

\begin{IEEEbiographynophoto}
	{\textsc{Mugen Peng}} [F] (pmg@bupt.edu.cn) is a Full Professor at the School of Information and Communication Engineering, Beijing University of Posts and Telecommunications, where he has served as the Dean of the School since June 2020 and as the Deputy Director of the State Key Laboratory of Networking and Switching Technology since October 2018.
	His research interests include wireless communication theory, radio signal processing, cooperative communication, self-organizing networks, LEO satellite communications, and the Internet of Things.
\end{IEEEbiographynophoto}

\begin{IEEEbiographynophoto}
	{\textsc{Yaohua Sun}} (sunyaohua@bupt.edu.cn) received his Ph.D. degree in communication engineering from Beijing University of Posts and Telecommunications (BUPT), Beijing, China, in 2019.
	He is currently an Associate Professor at the School of Information and Communication Engineering, BUPT.
	His research interests include intelligent radio access networks and LEO satellite communications.
\end{IEEEbiographynophoto}

\end{document}